\documentclass[10pt]{article}

\usepackage{fullpage}
\usepackage{setspace}
\usepackage{parskip}
\usepackage{titlesec}
\usepackage{xcolor}
\usepackage{lineno}
\usepackage[export]{adjustbox}

\usepackage[sorting=none, url=false, style=ieee]{biblatex}
\usepackage[version=4]{mhchem}
\usepackage{siunitx}
\DeclareSIUnit\Molar{M}

\usepackage{mathtools}

\PassOptionsToPackage{hyphens}{url}
\usepackage[colorlinks = true,
            linkcolor = blue,
            urlcolor  = blue,
            citecolor = blue,
            anchorcolor = blue]{hyperref}
\usepackage{etoolbox}

\renewenvironment{abstract}
  {{\bfseries\noindent{\abstractname}\par\nobreak}\normalsize}
  {\bigskip}

\titlespacing{\section}{0pt}{*3}{*1}
\titlespacing{\subsection}{0pt}{*2}{*0.5}
\titlespacing{\subsubsection}{0pt}{*1.5}{0pt}

\usepackage{authblk}

\usepackage{graphicx}
\usepackage[space]{grffile}
\usepackage{latexsym}
\usepackage{textcomp}
\usepackage{longtable}
\usepackage{tabulary}
\usepackage{booktabs,array,multirow}
\usepackage{amsfonts,amsmath,amssymb}
\usepackage{biblatex}

\AtBeginDocument{\DeclareGraphicsExtensions{.pdf,.PDF,.eps,.EPS,.png,.PNG,.tif,.TIF,.jpg,.JPG,.jpeg,.JPEG}}

\usepackage[utf8]{inputenc}
\usepackage[english]{babel}

\providecommand{\keywords}[1]{\textbf{\textit{Keywords ---}} #1}

\begin{document}

\doublespacing

\title{pyAvalanches: A Python Package for Analyzing Spatiotemporal
Propagation in Neuronal Avalanches}

\author[a,*]{Morgane Marzulli}
\author[b,*]{Marianna Angiolelli} 
\author[a]{Camilla Mannino}
\author[b]{Matteo Demuru}
\author[b,c]{Pierpaolo Sorrentino}
\author[a,+]{Marie-Constance Corsi}

\affil[a]{Sorbonne Université, Paris Brain Institute-ICM, Inria, Inserm, AP-HP, Hôpital de la Pitié Salpêtrière, Paris, France}
\affil[b]{Department of Medical, Motor and Wellness Sciences, University of Naples “Parthenope”, Naples, Italy}
\affil[c]{Institute of Systems Neuroscience, Aix-Marseille University, INSERM, UMR1106, Marseille, France}
\affil[*]{Authors contributed equally}
\affil[+]{Corresponding author: 
Marie-Constance Corsi \url{marie-constance.corsi@inria.fr}}

\vspace{-1em}
\begingroup
\let\center\flushleft
\let\endcenter\endflushleft
\maketitle
\endgroup

\selectlanguage{english}

\newpage
\begin{abstract}
The analysis of neuronal avalanches offers insights into brain dynamics utilizing the framework of criticality, but the reproducibility and comparability of studies are limited by the use of fragmented, lab-specific scripts. To address this issue, we introduce \texttt{pyAvalanches}, an open-source Python package providing a standardized, end-to-end pipeline for avalanche analysis from electrophysiological recordings (e.g., electroencephalography-EEG). Starting from the detection of neuronal avalanches the package provides their core statistical characterization, including size and duration distributions. Beyond this, the main aim of \texttt{pyAvalanches} is to characterize the spatiotemporal organization of activity propagation during avalanches. To this end, the core innovation of \texttt{pyAvalanches} is the compuation of Avalanche Transition Matrices (ATMs) to map spatiotemporal propagation patterns. Building on this, the package derives network-based metrics from the ATMs, bridging the study of the topology and organization of the underlying dynamical interactions with network neuroscience adopting the framework of neuronal avalanches. The entire workflow is encapsulated in a modular and scikit-learn compatible architecture. We demonstrate the utility of \texttt{pyAvalanches} through an illustrative group-level analysis on a public resting-state EEG dataset, comparing propagation patterns across different clinical populations. By providing a user-friendly, tested, and extensible tool, \texttt{pyAvalanches} facilitates reproducible research, enables the development of novel avalanche-based biomarkers, and makes complex avalanche analysis accessible to a broader scientific community. The package is fully documented and distributed via the Python Package Index (PyPI).
\end{abstract}

\newpage

\section{Introduction}

The study of neuronal avalanches, cascading bursts of neuronal activity  propagating through networks of brain cells, has provided new insights into brain dynamics and function \cite{cocchi_criticality_2017, shriki_neuronal_2013, rabuffo_neuronal_2021}. Each avalanche can be characterized by its size, typically defined as the total number of activation events occurring during the cascade, and its duration, corresponding to the time interval between its onset and termination. The approximately scale-free properties observed in avalanche size and duration distributions have been interpreted as evidence that brain dynamics may operate near criticality, a regime optimized for information processing, dynamic range, and computational power \cite{beggs_neuronal_2003,beggs_criticality_2008,cocchi_criticality_2017}.

Analyzing these distributions typically involves aggregating avalanche measurements across all system elements, which assumes uniform behavior across regions. While this homogeneity assumption may be valid for other physical systems exhibiting avalanche dynamics, it is difficult to justify in the brain, where regions differ markedly in structure and function.  Consequently, where an avalanche propagates and, in particular, the sequence in which brain regions are recruited, may carry physiologically and pathologically relevant information that cannot be captured by size and duration distributions alone.

\begin{figure}
	\centering
	\includegraphics[width=1\columnwidth]{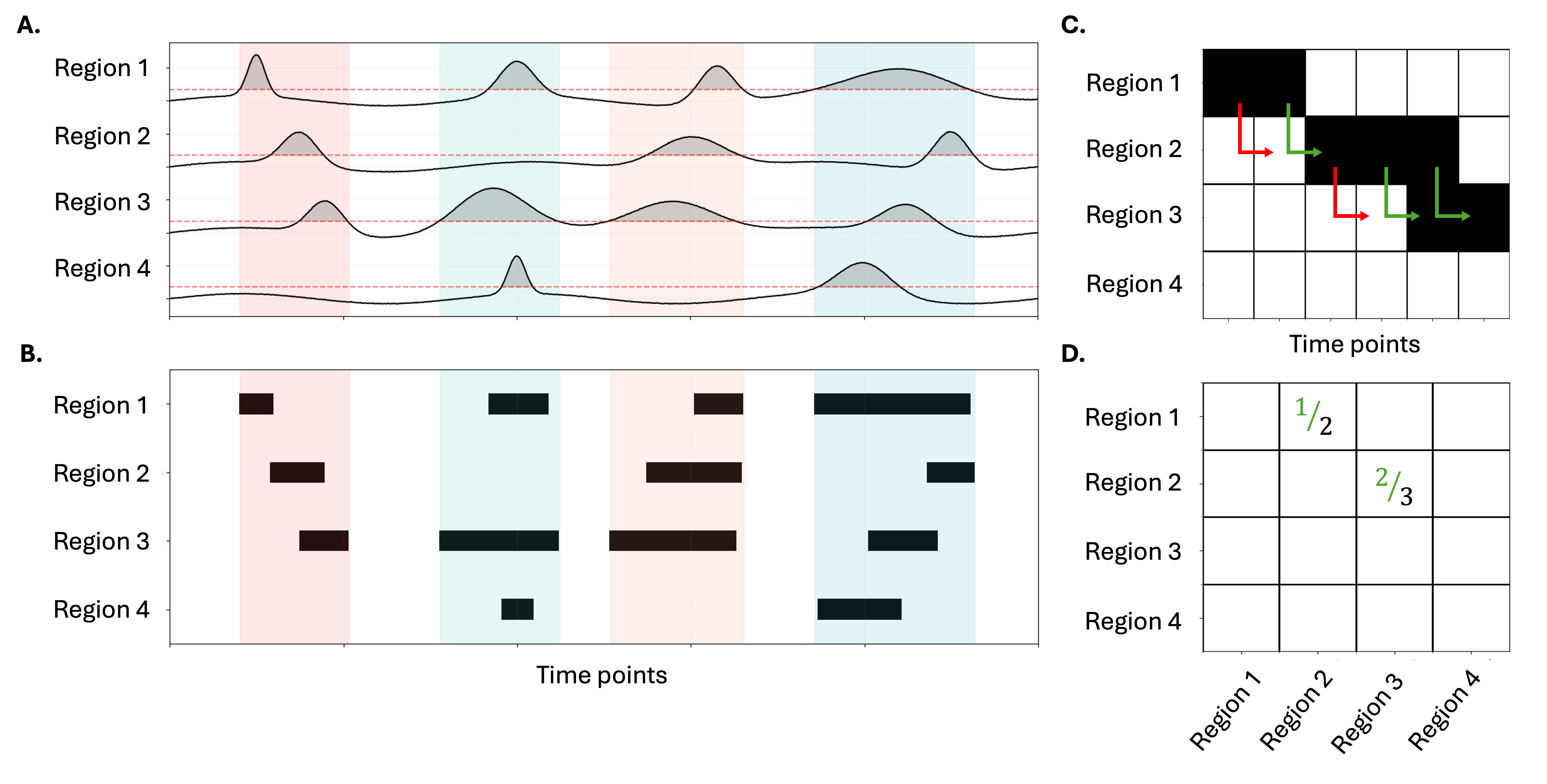}
	\caption{Conceptual overview of avalanches and Avalanche Transition Matrices. (A) Simulated multichannel EEG signal. Each line represents the z-scored signal obtained from a specific brain region. Red dashed horizontal lines indicate the activation threshold for each region; background shading highlights time windows with at least one active signal. (B) Corresponding raster plot where horizontal lines mark suprathreshold activity for each channel. The colored background regions spanning the full height of the raster plot delineate distinct neural avalanches - periods of cascading activity where at least one region exhibits suprathreshold events. (C) Avalanche transition matrix estimation procedure. During a representative avalanche, Region 1 is active twice and Region 2 is active three times. Green arrows indicate a transitions (e.g., when Region 1 is active, Region 2 subsequently activates). Red arrows indicate no transitions (e.g., when Region 1 is active but Region 2 does not activate). The same logic is applied to track transitions from Region 2 to Region 3. (D) Transition probability matrix derived from the avalanche observations in panel C. Matrix entries represent the conditional probability that a target region activates given that a source region was active in the preceding time step. For example, the probability of Region 2 activating given Region 1 was active is 1/2, while the probability of Region 3 activating given Region 2 was active is 2/3. Adapted from Sorrentino et al. \cite{sorrentino_structural_2021}.}
	\label{FIG:1}
\end{figure}

To characterize this spatiotemporal organization, the Avalanche Transition Matrix (ATM) was developed to quantify the propagation of neuronal avalanches between brain regions \cite{sorrentino_structural_2021}. By assessing the probability of transitions between active brain regions during an avalanche, the ATM links large-scale neuronal dynamics to the organization of underlying neural circuits, thereby providing a directed representation of the preferential pathways through which avalanche activity unfolds across the brain.
An ATM is a matrix $A$ where each element $a_{ij}$ represents the probability that, during avalanches, one region \textit{j} activates following the activation of region \textit{i} (Fig.\ref{FIG:1}). 

Previous studies shown its potential in applications ranging from Brain-Computer Interfaces (BCI) \cite{corsi_measuring_2024, mannino_neuronal_2025}, to large-scale clinical studies \cite{romano_topological_2023, duma_altered_2024, corsi_neuronal_2024, polverino_altered_2024, agouram_l-dopa-induced_2025, mannino_weighted-stochastic_2026}, to the characterization of healthy brain dynamics \cite{sorrentino_brain_2023}.

Nonetheless, the study of neuronal avalanches, especially their spatiotemporal dynamics, lacks unified analysis toolboxes. Research has largely relied on custom, lab-specific scripts, leading to a fragmented landscape where reproducibility is challenging. The sensitivity of results to arbitrary choices in event detection, thresholding, binning often yields non-comparable findings across studies. This creates a need for an open-source, standardized, and user-friendly tool that can provide a consistent framework for this complex analysis.

While packages exist for the statistical fitting of power-law distributions (e.g., powerlaw \cite{alstott_powerlaw_2014}), they do not handle the step of event extraction from raw neurophysiological signals. Conversely, dedicated toolboxes like \texttt{CROCOpy} \cite{myrov_crocopy_2026}; \texttt{EDGE OF PY} \cite{obyrne_jnobyrneedgeofpy_2026}; or \texttt{CASCADE} \cite{avila_cascade_2026} focus primarily on avalanche detection and criticality metrics but lack integrated tools for analyzing their spatiotemporal propagation. To the best of our knowledge, no toolboxes have currently implemented ATMs computation.

To address these limitations, we introduce \texttt{pyAvalanches}, an open-source Python package that provides a complete, end-to-end pipeline for detecting, characterizing, and analyzing neuronal avalanches from neurophysiological recordings (e.g., electroencephalography (EEG), magnetoencephalography (MEG), electrocorticography (ECoG)). \texttt{pyAvalanches} guides users from time-series data through avalanche detection, spatiotemporal characterization via ATMs, and extraction of network-based features. It is the first package to integrate the computation of ATMs and their derived network metrics directly into the analysis workflow.

The core contribution of \texttt{pyAvalanches} is to provide a unified framework that combines traditional avalanche analyses (size and duration distributions) with advanced spatiotemporal pattern analyses. It is compatible with scikit-learn \cite{pedregosa_scikit-learn_2011}, enabling the direct use of ATM features in classification and other predictive models. Furthermore, it includes automated .csv files and report generation to facilitate rapid results exploration. It requires very little coding or signal processing experience, making it suitable for a very large scientific audience. The package is fully documented and distributed via the Python Package Index (PyPI).

\section{Software details}
\subsection{Core functionalities}

\texttt{pyAvalanches} is built around four main core functionalities:

\begin{enumerate}

\item 
\textit{Avalanche detection}: identifies neuronal avalanches from multi-channel time-series data (e.g., EEG, MEG, ECoG). The detection pipeline is highly configurable, with parameters for event thresholding (e.g., standard deviation, absolute value), temporal binning, and data normalization. It computes a wide range of avalanche properties, including size, duration, and spatiotemporal patterns, and allows the visualization of power-law fitting to support the parametrization of the pipeline.

\item
\textit{Avalanches Transition Matrices computation}:  characterizes the propagation of activity during avalanches. The package implements two distinct methods for ATM computation: the standard conditional probability model \cite{sorrentino_structural_2021} and the weighted-stochastic formulation that incorporates the length of regional activation periods and avalanche sizes into the transition probabilities \cite{mannino_weighted-stochastic_2026}.

\item
\textit{Network analysis}: Computes graph-theoretic metrics directly from the generated ATMs, bridging avalanche-dynamics analysis with established principles of network neuroscience. This functionality leverages specialized libraries such as the Brain Connectivity Toolbox \cite{aestrivex_aestrivexbctpy_2026} and NetworkX \cite{hagberg_exploring_2008} to quantify network properties such as node centrality and strength.

\item 
\textit{Results management and reporting}: provides dedicated data containers to systematically organize results and metadata at both the subject and group levels. This structure streamlines data handling and simplifies the export of features for subsequent population-level statistical analysis. The package also includes a function for the automated generation of comprehensive HTML reports and utility functions for creating summary plots and pandas DataFrames.

\end{enumerate}

\subsection{Software architecture}

\texttt{pyAvalanches} is implemented with a modular, sequential pipeline architecture. The core workflow is divided into distinct stages for each functionality described above (avalanche detection, ATM computation, and metrics extraction) each encapsulated in its own pipeline object. The output of each stage is a dedicated results container that serves as the input for the next, creating a clear and intuitive data flow. This modular design provides flexibility, allowing users to execute only the required portions of the analysis (e.g., performing only avalanche detection without computing ATMs). 

The software is built upon the standard Python scientific computing stack, with key dependencies including NumPy \cite{harris_array_2020}, SciPy \cite{virtanen_scipy_2020}, and matplotlib \cite{hunter_matplotlib_2007}. Furthermore, the ATM pipeline includes scikit-learn \cite{pedregosa_scikit-learn_2011} compatible transformers, allowing the resulting matrices to be directly extracted and vectorized as features that can be fed into machine learning classifiers.

To ensure broad compatibility, \texttt{pyAvalanches} supports both continuous (resting-state) and epoched (task-related) data, which can be provided as NumPy arrays or as MNE-Python \cite{gramfort_meg_2013} objects, allowing it to work with well established signal processing pipelines.

\subsection{Class and methods structure}

\begin{figure*}
	\centering
	\includegraphics[width=1\textwidth]{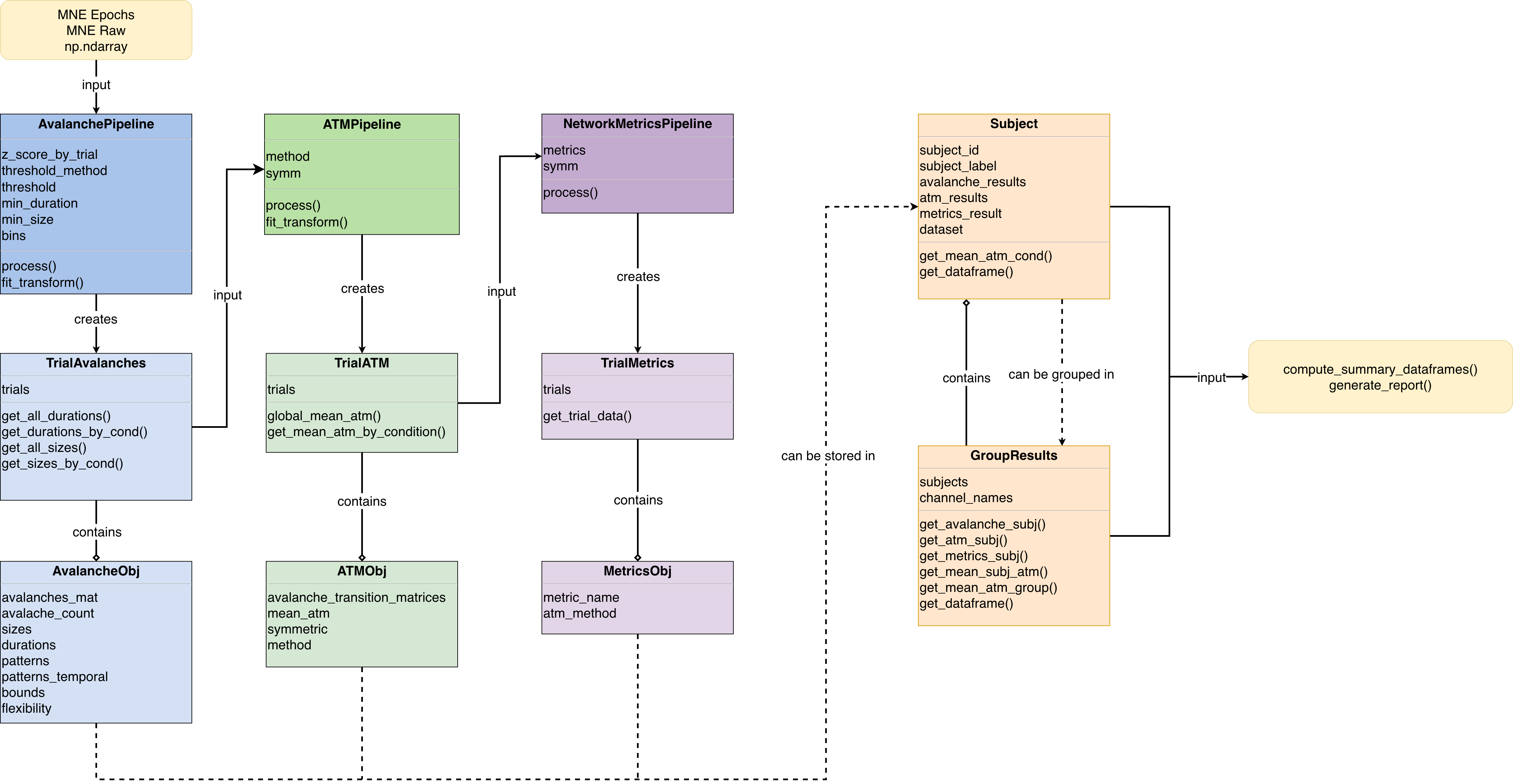}
	\caption{The \texttt{pyAvalanches} class architecture. The software is structured into pipeline classes (e.g., \texttt{AvalanchePipeline}) that execute the analysis and container classes (e.g., \texttt{TrialAvalanches}) that store results. The arrows indicate the composition relationships and the sequential workflow with final aggregation in the \texttt{Subject} and \texttt{GroupResults} objects.}
	\label{FIG:2}
\end{figure*}

\texttt{pyAvalanches} is centered on two main class types: pipelines for processing and containers for results (Fig.\ref{FIG:2}).

Each step of the analysis is performed by its dedicated pipeline (\texttt{AvalanchePipeline}, \texttt{ATMPipeline} and \texttt{NetworkMetricsPipeline}), each with a consistent \texttt{.process()} method that performs the computations. The outputs of the pipelines are stored in corresponding container classes (\texttt{TrialAvalanches},\texttt{TrialATM} and \texttt{TrialMetrics}) which organize data on a per-trial (i.e. contiguous segment of neural data) basis. These containers provide useful methods for aggregation and filtering by experimental condition (e.g., \texttt{.get\_all\_durations()}, \texttt{.global\_mean\_atm()}).

For population-level studies, the \texttt{Subject} class aggregates all results for a single participant, and the \texttt{GroupResults} class manages collections of \texttt{Subject} objects, both providing methods to export structured results directly into pandas DataFrames for subsequent statistical analysis.

\section{Illustrative example}

\begin{figure*}
	\centering
	\includegraphics[width=1
    \textwidth]{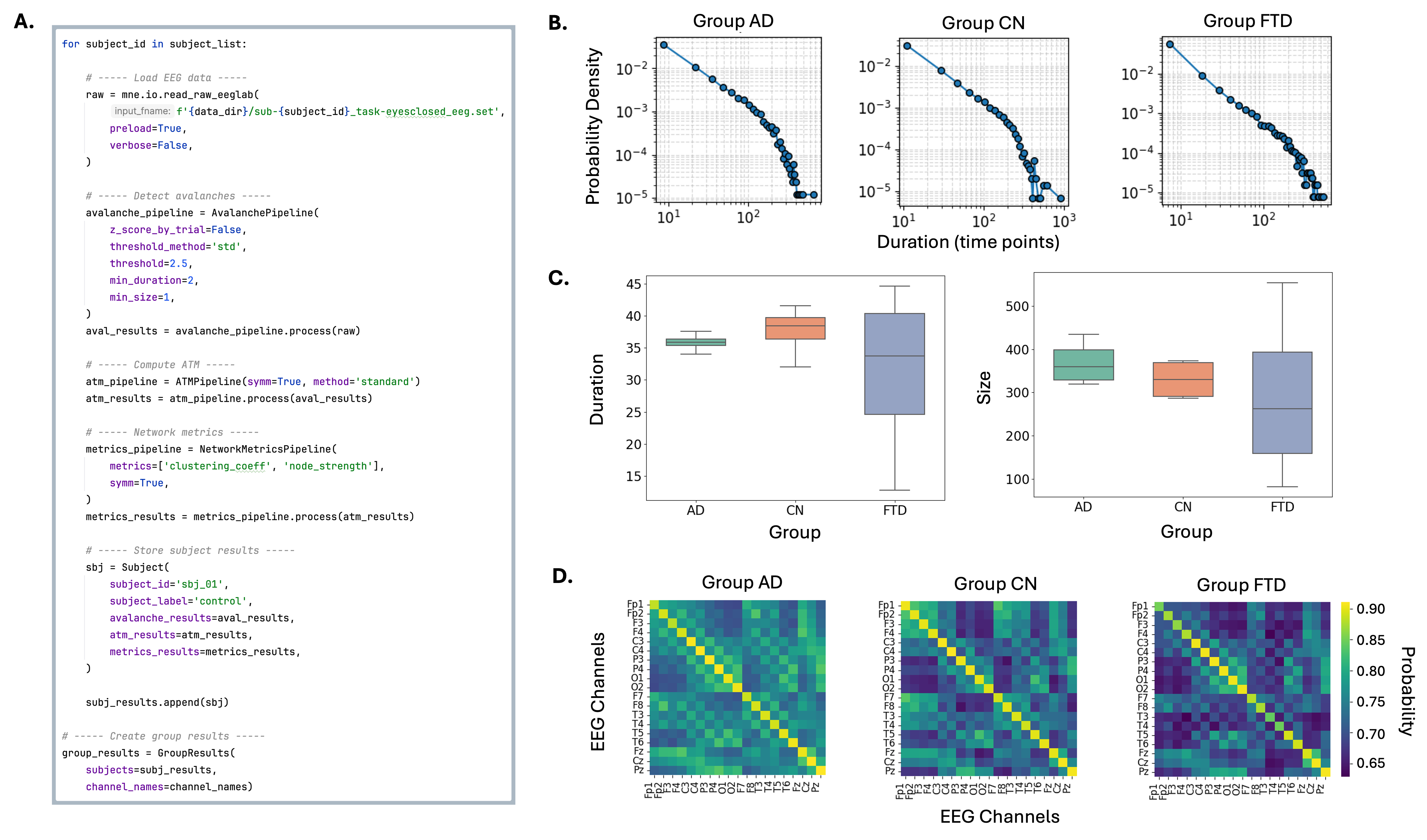}
	\caption{End-to-end group-level analysis workflow and key outputs with \texttt{pyAvalanches}. The workflow is demonstrated on a resting-state EEG dataset with three clinical groups: Alzheimer's Disease (AD), Frontotemporal Dementia (FTD), and healthy controls (CN). (A) A simplified representation of the core Python script, showing the sequential application of pipelines to each subject's data and the final aggregation into a \texttt{GroupResults} object. (B) Group-level avalanche duration distributions plotted on a log-log scale, showing the power-law distribution for each condition. (C)  Box plots of the avalanche durations and size distributions for each clinical group. (D) Mean Avalanche Transition Matrix (ATM) for each group, averaged across subjects.}
	\label{FIG:3}
\end{figure*}

To demonstrate a typical end-to-end workflow, we present a group-level analysis of a publicly available resting-state EEG dataset \cite{miltiadous_dataset_2026}, containing recordings from patients with Alzheimer’s Disease (AD), Frontotemporal Dementia (FTD), and healthy controls (CN). The complete analysis, summarized in Fig.\ref{FIG:3}, follows a three-stage process: per-subject processing, group-level aggregation, and result visualization.

The core code, shown in Fig.\ref{FIG:3}A, iterates through each subject's data. For each individual, the workflow loads the EEG data using MNE-Python, then applies a sequence of \texttt{pyAvalanches} pipelines to detect avalanches (\texttt{AvalanchePipeline}), compute the related ATMs (\texttt{ATMPipeline}), and extract network metrics (\texttt{NetworkMetricsPipeline}). The results from all stages are encapsulated within a \texttt{Subject} object. Once all individuals are processed, the list of \texttt{Subject} objects is aggregated into a single \texttt{GroupResults} container, which serves as the entry point for all subsequent group-level analyses.

This container is then used to generate key summary outputs. The distributions of avalanche durations and sizes are plotted for each clinical group (Fig.\ref{FIG:3}B and Fig.\ref{FIG:3}C). The primary output, the group-averaged ATM for each condition, is visualized as a heatmap, allowing for direct comparison of large-scale spatiotemporal propagation patterns (Fig.\ref{FIG:3}D). This example highlights how \texttt{pyAvalanches} provides an easy and straightforward framework for conducting complex group-level analyses.

\section{Impact and future directions}

\texttt{pyAvalanches} is designed to advance the study of neuronal avalanche dynamics and their spatiotemporal organization. 
By offering a standardized and open-source platform, it facilitates benchmarking and the direct comparison of results across studies, promoting the FAIR principles and making these complex analyses accessible to a wider scientific audience. 
While neuronal avalanches have historically played an important role in the investigation of brain criticality, \texttt{pyAvalanches} is not restricted to testing criticality hypotheses. Rather, it provides a general framework for detecting and characterizing avalanche dynamics and for investigating how neural activity propagates across brain networks. Furthermore, the package creates a new bridge between the study of neuronal avalanches and network neuroscience: through the integrated computation of ATMs and the subsequent application of graph-theoretic metrics, \texttt{pyAvalanches} allows researchers to identify functional hubs and quantify the topological properties of the dynamic networks through which activity propagates. Finally, the package's integration with scikit-learn allows the use of new ATM derived features in predictive models for tasks such as classifying clinical populations, tracking disease progression, or decoding cognitive states.

Building on this extensible foundation, future development will focus on adding new metrics, for both avalanche properties and network topology. The future goals include also the integration of information from neuroimaging data, enabling researchers to combine the analysis of fast electrophysiological dynamics with data from structural (e.g., diffusion tensor imaging - DTI) or functional (e.g., functional magnetic resonance imaging - fMRI) imaging.

\section{Code Availability}
\label{sec:availability}

\texttt{pyAvalanches} is an open-source Python package released under the BSD-3-Clause license. The source code is
hosted on GitHub at \url{https://github.com/morgane-marz/pyAvalanches}, where users can
report issues, request features, and contribute via pull requests.

\paragraph{Installation.}
\texttt{pyAvalanches} requires Python $\geq 3.9$ and can be installed from
the Test Python Package Index (TestPyPI):
\begin{verbatim}
pip install -i https://test.pypi.org/simple/ pyAvalanches
\end{verbatim}

\paragraph{Documentation.}
Comprehensive documentation, including installation instructions, tutorials,
example notebooks, and a full API reference, is available at
\url{https://pyavalanches.readthedocs.io/en/latest/}.

\section{Acknowledgments}
This work was supported by the French National Research Agency (MANET project, ANR-25-CE33-7747-01), the IRIN (Italian Research Infrastructure for Neuroscience, Project “EBRAINS-Italy”, Next GenerationEU/Italian National Recovery and Resilience Plan (NRRP), M4C2 (Project code IR0000011, CUP B51E22000150006) and Governo Italiano Ministero per lo sviluppo Economico (ACCORDI PER INNOVAZIONE. Approccio User-friendly integrato per Diagnosi, Assistenza e Cura Efficaci—AUDACE grant number B69J23006050007). The funders had no role in study design, data collection and analysis, decision to publish, or preparation of the manuscript.


\printbibliography

\selectlanguage{english}
\end{document}